\documentclass[12pt]{article}
\pdfoutput=1
\usepackage{geometry,enumerate,amsmath,amssymb}
\usepackage{fullpage}
\usepackage{graphicx}
\usepackage{bm}% bold math
\numberwithin{equation}{section}
\newcommand{\be}{\begin{equation}}
\newcommand{\ee}{\end{equation}}
\newcommand{\bea}{\begin{eqnarray}}
\newcommand{\eea}{\end{eqnarray}}
\renewcommand{\epsilon}{\varepsilon}
\newcommand{\bx}{\boldsymbol{x}}

\begin{document}
\title{Rational Skyrmions}
\author{Derek Harland$^\star$ and Paul Sutcliffe$^\dagger$\\[20pt]
{\em \normalsize $^\star$ 
School of Mathematics, University of Leeds, Leeds LS2 9JT, UK,}\\
{\normalsize Email: d.g.harland@leeds.ac.uk}\\[5pt]
{\em \normalsize $^\dagger$ 
Department of Mathematical Sciences,}\\
{\em \normalsize Durham University, Durham DH1 3LE, U.K.}\\ 
{\normalsize Email: p.m.sutcliffe@durham.ac.uk}\\ 
}
 \date{August 2023}

\maketitle
\begin{abstract}
  A new method is introduced to construct approximations to Skyrmions that are explicit rational functions of the spatial Cartesian coordinates. The scheme uses ADHM data of a Yang-Mills instanton to produce a Skyrmion with a baryon number that is equal to the instanton number. The formula for the Skyrmion involves only the evaluation of the ADHM data, in contrast to the Atiyah-Manton construction that requires the solution of a differential equation that can only be solved explicitly in the case of a spherically symmetric Skyrmion. Examples with baryon numbers one and two are studied in detail. The energy of the rational Skyrmion with baryon number one is lower than that of the Atiyah-Manton Skyrmion, which is already within one percent of the energy of the true numerically computed Skyrmion. A family of baryon number two Skyrmions is presented, which includes an axially symmetric Skyrmion that smoothly transforms to a pair of well-separated single Skyrmions as the parameter is varied.  
    \end{abstract}

\newpage
\section{Introduction}\quad
Skyrmions are topological solitons that model baryons within a nonlinear theory of pions, where the integer-valued topological charge is identified with baryon number \cite{Sk,Manton}. Static Skyrmions are obtained by computing solutions of a nonlinear partial differential equation for the $SU(2)$-valued Skyrme field, or equivalently by minimizing the static energy within a given topological sector. This requires numerical methods, as there are no explicit Skyrmion solutions of the Skyrme model. For baryon number one, spherical symmetry may be imposed to reduce the partial differential equation to an ordinary differential equation for a radial profile function, but even this profile function must be computed numerically. This lack of explicit Skyrmions has motivated several methods to obtain Skyrme fields that give reasonable approximations to Skyrmions.

The rational map approximation \cite{HMS} is a kind of separation of variables, where the angular dependence of the Skyrme field is specified explicitly in terms of a rational map between Riemann spheres. This is a semi-analytic method, as the radial dependence is via a profile function that is computed numerically to minimize the energy. For baryon number one the rational map approximation is simply a reformulation of the imposition of spherical symmetry. For baryon numbers greater than one it produces non-spherical fields that, for a suitable choice of rational map, match the symmetries of the true minimal energy Skyrmions and have energies that are accurate to within a few percent \cite{BS}, at least in the case of massless pions that is the situation under consideration in the present paper.

The main disadvantage of the rational map approximation is that it doesn't allow the minimal energy Skyrmions to separate into individual Skyrmions, or lower charge clusters. It is important to have a good description of this situation to understand the interactions between Skyrmions and their scattering processes. Moreover, it is necessary for providing a suitable moduli space for quantization that allows more degrees of freedom than simply translations, rotations and isorotations, as it is known that restricting to such zero mode quantization is inadequate.

Well-separated Skyrmions can be approximated by a product ansatz, where the Skyrme fields of the individual Skyrmions are simply multiplied together. However, the product ansatz fails to provide a good description of Skyrmions once they are no longer well-separated, and in particular is not suitable to approximate minimal energy Skyrmions with baryon numbers greater than one. This method is therefore complementary to the rational map approximation, but unfortunately no way is known to patch these two techniques together.

The only method that is currently capable of describing the whole gamut of Skyrmions, from a collection of well-separated single Skyrmions all the way through to the minimal energy Skyrmions formed as they merge, is the Atiyah-Manton instanton holonomy procedure \cite{AM,AM2}. The starting point for this approach is self-dual Yang-Mills instantons in four-dimensional space. This is an integrable conformal theory with instanton moduli spaces that can be obtained using the ADHM construction \cite{ADHM}. The Atiyah-Manton Skyrme field is equal to the holonomy of the instanton along lines parallel to the extra dimension, and has a baryon number that is equal to the instanton number.  For a suitably chosen instanton the Skyrme field provides a good approximation to the minimal energy Skyrmion for any baryon number. Furthermore, by varying the instanton moduli the minimal energy Skyrmion can be separated into its single Skyrmion constituents, thereby providing a Skyrmion moduli space induced from the instanton moduli space. In the case of baryon number one, the Atiyah-Manton Skyrmion is spherically symmetric with an explicit profile function that depends on the arbitrary scale of the instanton. For a particular value of the instanton scale the energy is within $1\%$ of the true Skyrmion energy.

A disadvantage of the Atiyah-Manton prescription is that the calculation of the holonomy requires the solution of a family of ordinary differential equations, that can only be solved explicitly in the case of a spherically symmetric Skyrmion. This means that the computation of minimal energy  Atiyah-Manton Skyrmions with baryon numbers greater than one must be implemented numerically. Such numerical computations reveal that the approximations are very good, with an energy excess compared to the true numerically calculated Skyrmions that is typically less than $2\%$. The success of the Atiyah-Manton construction can be explained  by an exact equivalence between the four-dimensional Yang-Mills theory and the three-dimensional Skyrme model coupled to an infinite tower of vector mesons \cite{Su}. The Atiyah-Manton approximation appears within this framework as the leading order term in a basis expansion when the vector mesons are neglected. 

Recently a new numerical method has been introduced \cite{CHW} and developed  \cite{Ha} to compute the instanton holonomy required for the construction of the Atiyah-Manton Skyrmion. The method is geometrically natural and is based on the theory of parallel transport of an induced connection. Not only is the scheme more efficient than the standard approach, based on a traditional differential equation solver, but it removes all issues associated with gauge variations. In fact, this is a suite of schemes with increasing order to decrease the numerical errors that appear in any discretization that replaces the line with a finite set of points.

The main idea that lies at the heart of the work in the present paper may at first sight seem bizarre. The proposal is to use the least accurate of the new numerical schemes with an extremely coarse discretization. In fact, it is an ultra-discrete approach in which the real line (along which the holonomy is calculated) is replaced by only five points, $\pm \infty,\pm\mu,0$, where $\mu$ is a positive constant. However, the geometry behind the computational scheme saves the day by preserving the topology, in the sense that the topological charge of the Skyrme field remains equal to the instanton number.

The advantage of this ultra-discrete method is that it turns the numerical scheme into an analytic approach. Skyrme fields are obtained in explicit closed form as rational functions of the spatial Cartesian coordinates by simply evaluating the ADHM data of the instanton. This is an analytic alternative to the Atiyah-Manton method that avoids the need to solve any differential equations and instead uses a simple formula for the rational Skyrmion in terms of the ADHM data. Examples for charges one and two are presented in detail to confirm that the approach does indeed yield good approximations. In particular, for charge one the rational Skyrmion has an energy that is even lower than the Atiyah-Manton Skyrmion.

\section{Skyrmions and instantons}\quad
A Skyrme field $U(\bx)$ is a smooth map from $\mathbb{R}^3$ to $SU(2)$ that satisfies the boundary condition $U\to 1$ as $|\bx|\to\infty.$ Maps of this form have an associated topological charge, $B\in\mathbb{Z}=\pi_3(SU(2))$, that may be computed
as
\be
B=\int \frac{1}{24\pi^2}\varepsilon_{ijk}\mbox{Tr}(R_iR_kR_j)\, d^3x,
\label{baryon}
\ee
where the $\mathfrak{su}(2)$-valued right currents are $R_i=\partial_iU\,U^{-1}.$ Physically, the charge $B$ corresponds to baryon number.

In suitable units, the static Skyrme energy is 
\be
E=\frac{1}{12\pi^2}\int 
-\mbox{Tr}\bigg\{\frac{1}{2}R_i^2+\frac{1}{16}[R_i,R_j]^2
\bigg\}\,d^3x,
\label{energy}
\ee
and obeys the Faddeev-Bogomolny energy bound \cite{Fa}, which states that
$E\ge B$. The only Skyrme field that attains this bound is the trivial vacuum solution $U=1$ with $B=0.$

The study of Skyrmions begins with finding the minimal energy charge $B$ Skyrmion, that is, the Skyrme field $U(\bx)$ that minimizes the energy (\ref{energy}) within the sector with topological charge equal to $B$. For $B=1$ the minimal energy Skyrmion is spherically symmetric, taking the hedgehog form
\be
U=\cos f +i \frac{\sin f}{r}\, {\bx \cdot \boldsymbol{\tau}},
\label{hedgehog}
\ee
where $\boldsymbol{\tau}$ denotes the triplet of Pauli matrices,
$r^2=|\bx|^2=x_1^2+x_2^2+x_3^2$, and
$f(r)$ is a real radial profile function satisfying the boundary conditions
$f(0)=\pi$ and $f(\infty)=0.$ Substituting (\ref{hedgehog}) into (\ref{energy}) reveals that $f(r)$ is determined by minimizing the energy
\be
E=\frac{1}{3\pi}\int_0^\infty \bigg(r^2f'^2+2(1+f'^2)\sin^2 f+\frac{\sin^4 f}{r^2}\bigg) \, dr.
\label{sphericalenergy}
\ee
The energy minimizing profile function can only be computed numerically and gives the value $E=1.232$ for a single Skyrmion. For $B>1$ the minimal energy Skyrmion is not spherically symmetric, but it is a bound state, as it has an energy that is less than $B$ times the energy of the charge one Skyrmion.

The Atiyah-Manton construction  \cite{AM,AM2} of Skyrme fields from instantons  begins with the gauge potential $A_\mu(\bx,x_4)\in \mathfrak{su}(2)$ of a self-dual Yang-Mills instanton in $\mathbb{R}^4$, with instanton number $N$. The prescription for the Skyrme field is to take the holonomy of the instanton along the $x_4$ direction,
$U(\bx)=\Omega(\bx ,\infty)$, where $\Omega(\bx,x_4)$ is the solution of the matrix ordinary differential equation
\be
\partial_4 \Omega(\bx,x_4)=\Omega(\bx,x_4)A_4(\bx,x_4),
\label{holonomy}
\ee
with initial condition $\Omega(\bx,-\infty)=1.$ The resulting Skyrme field has a topological charge $B$ that is equal to the instanton number $N$.

There is an $8N$-dimensional moduli space of charge $N$ instantons. In the part of the moduli space that describes well-separated single instantons, these moduli have the interpretation of a position in $\mathbb{R}^4$, a scale, and an $SU(2)$ orientation, for each individual instanton.  For each $N$, a suitable point in the moduli space generates a good approximation to the minimal energy Skyrmion with baryon number $B=N.$ For $N=1$, taking an instanton at the origin with scale $\lambda$ produces an Atiyah-Manton Skyrmion of the hedgehog form (\ref{hedgehog}) with a profile function
\be
f(r)=\pi\bigg(1-\frac{r}{\sqrt{\lambda^2+r^2}}\bigg).
\label{amprofile}
\ee
The optimal energy minimizing instanton scale is $\lambda=1.45,$
with an energy $E=1.243$ that is less than $1\%$ above the true Skyrmion energy.

The $8N$-dimensional moduli space of charge $N$ instantons can be obtained via the ADHM construction \cite{ADHM}, which provides an equivalence between the instanton moduli spaces and certain quaternionic matrices, known as ADHM data. This data consists of a matrix
\be
\widehat M = \begin{pmatrix} L\\ M \end{pmatrix},
\ee
where $L$ is a row of $N$ quaternions and $M$ is a symmetric $N\times N$ matrix of quaternions. ADHM data must satisfy the condition that
the $N\times N$ matrix $\widehat M^\dagger \widehat M$
is real and non-singular, where ${}^\dagger$ denotes the quaternionic conjugate transpose. The instanton gauge potential can be obtained from the ADHM data using quaternionic linear algebra to calculate the kernel of the operator
\be
\Delta(\bx,x_4)=\begin{pmatrix}L\\ M-(ix_1+jx_2+k x_3+x_4) 1_{N}\end{pmatrix},
\label{adhmoperator}
\ee
where $1_N$ denotes the $N\times N$ identity matrix.
Explicitly, let $\Psi(\bx,x_4)$ be an $(N+1)$-component column vector of unit length, $\Psi(\bx,x_4)^\dagger\Psi(\bx,x_4)=1$, that solves the equation
\be
\Psi(\bx,x_4)^\dagger\Delta(\bx,x_4)=0.
\ee
The self-dual gauge potential of the instanton is given by
\be
A_\mu=\Psi(\bx,x_4)^\dagger\partial_\mu\Psi(\bx,x_4),
\label{potential}
\ee
which is a pure quaternion that can be identified with an element of $\mathfrak{su}(2)$ in the standard way, by relating quaternions to Pauli matrices.

Unfortunately, even when the ADHM construction can be used to obtain an explicit instanton gauge potential $A_\mu$, the ordinary differential equation (\ref{holonomy}) cannot be solved explicitly in closed form. The only exception to this statement is the case that the holonomy produces a spherically symmetric Skyrme field of the hedgehog form (\ref{hedgehog}). In all other cases the holonomy must be computed numerically, including all examples for approximations to the minimal energy Skyrmions with $B>1.$

\section{Rational Skyrmions from ADHM data}\quad
New computational schemes have recently been introduced \cite{Ha} to calculate the holonomy $\Omega(\bx,\infty)$ of an instanton, given the associated ADHM data.
These are finite difference methods, of various orders, that all involve replacing the real line, parameterized by $x_4$, with a set of $p$ points $-\infty=t_1<t_2<...<t_p=\infty$. The scheme of interest here is the lowest order method \cite{CHW}, where derivatives are replaced by first-order forward difference approximations.

The starting point to derive the lowest-order scheme is to substitute the formula (\ref{potential}) for the gauge potential into the
holonomy equation (\ref{holonomy}) to give
\be
\partial_4 \Omega(\bx,x_4)=\Omega(\bx,x_4)\Psi(\bx,x_4)^\dagger\partial_4\Psi(\bx,x_4).
\ee
Applying a forward difference approximation at the point $x_4=t_i$ to both derivatives in this equation, and neglecting the error terms, gives
\be
\Omega(\bx,t_{i+1})-\Omega(\bx,t_i)
=\Omega(\bx,t_i)\Psi(\bx,t_i)^\dagger(\Psi(\bx,t_{i+1})-\Psi(\bx,t_{i})).
\ee
Using the fact that $\Psi$ has unit length, this becomes the simple relation
\be
\Omega(\bx,t_{i+1})
=\Omega(\bx,t_i)\Psi(\bx,t_i)^\dagger\Psi(\bx,t_{i+1}),
\ee
with solution
\be
\Omega(\bx,\infty)=\Omega(\bx,t_p)=
\Psi(\bx,t_1)^\dagger\Psi(\bx,t_2)
\Psi(\bx,t_2)^\dagger\Psi(\bx,t_3)...\Psi(\bx,t_{p-1})^\dagger\Psi(\bx,t_p),
\label{dhol}
\ee
given that the initial condition is
$\Omega(\bx,t_1)=\Omega(\bx,-\infty)=1.$ Furthermore, as $|t_1|=|t_p|=\infty$, then the remaining freedom may be fixed by setting $\Psi(\bx,t_1)=\Psi(\bx,t_p)=e_1$, where $e_1$ is the $(N+1)$-component column vector with first entry equal to 1 and all other entries equal to $0$.

A simplification that was overlooked in \cite{Ha} is that
\be
\Psi(\bx,x_4)\Psi(\bx,x_4)^\dagger=Q(\bx,x_4),
\label{defQ}
\ee
where 
$Q(\bx,x_4)$ is the projector onto the kernel of $\Delta(\bx,x_4)$ that is
given by the $(N+1)\times (N+1)$ quaternionic matrix
\be
Q(\bx,x_4)=1_{N+1} -\Delta(\bx,x_4)\bigg(\Delta(\bx,x_4)^\dagger\Delta(\bx,x_4)\bigg)^{-1}\Delta(\bx,x_4)^\dagger,
\ee
and the ADHM construction requires that $\Delta(\bx,x_4)^\dagger\Delta(\bx,x_4)$ is a real non-singular matrix.
Substituting (\ref{defQ}) into (\ref{dhol}) allows the discrete holonomy to be calculated by simply evaluating the ADHM data at the $p-2$ interior lattice points
\be
\Omega(\bx,\infty)=e_1^\dagger Q(\bx,t_2)Q(\bx,t_3)...Q(\bx,t_{p-1})e_1.
\label{anyp}
\ee
The errors in the finite difference approximation mean that $\Omega(\bx,\infty)$ will not be a unit quaternion, therefore to identify the discrete holonomy with a Skyrme field $U(\bx)$ requires a final normalization step,
\be
U(\bx)=\frac{\Omega(\bx,\infty)}{|\Omega(\bx,\infty)|},
\ee
to produce the unit quaternion $U(\bx)$, that is identified with an element of $SU(2)$ by relating quaternions to Pauli matrices.

This computational scheme will now be converted into an analytic method by using an ultra-discrete version with only 5 lattice points, $t_1,..,t_5=-\infty,-\mu,0,\mu,\infty$, where $\mu$ is a positive parameter. The Skyrme field is therefore obtained by evaluating the ADHM data at only 3 interior points, $0,\pm\mu$, and is given explicitly by the formula
\be
U(\bx)
=\frac{e_1^\dagger Q(\bx,-\mu)Q(\bx,0)Q(\bx,\mu)e_1}
{|e_1^\dagger Q(\bx,-\mu)Q(\bx,0)Q(\bx,\mu)e_1|}.
\label{scheme5}
\ee
It would appear that (\ref{scheme5}) gives a Skyrmion that is an algebraic function of the Cartesian coordinates, rather than a rational function, because of the modulus in the denominator. However, in all the examples considered here, it turns out that the Skyrmion is rational.
The explanation for this surprising result is that (almost) all of the ADHM data investigated have the property that
\begin{equation}
kM=-Mk \quad \text{and}\quad kL=Lk. \label{axial}
\end{equation}
Equivalently, $M=M_1i+M_2j$ and $L=L_3k+L_4$ for real matrices $M_1,M_2,L_3,L_4$.  This is also equivalent to the instanton having an axial symmetry in the $(x_3,x_4)$-plane.  Axially-symmetric instantons correspond to hyperbolic monopoles \cite{At}, so this constraint is natural and not too restrictive.

The proof that (\ref{axial}) is a sufficient condition for the Skyrme field $U$ to be rational proceeds by first establishing the identity.
\be
  e_1^\dagger\big( Q(\bx,0)Q(\bx,\mu)-Q(\bx,-\mu)Q(\bx,0)\big)e_1 = 0
  \label{property}.
\ee
To prove this, first note that (\ref{axial}) implies that
\be
\Delta(\bx,\mu)^\dagger\Delta(\bx,\mu) = \Delta(\bx,0)^\dagger\Delta(\bx,0) + \mu^2 1_N = \Delta(\bx,-\mu)^\dagger\Delta(\bx,-\mu).
\ee
Using this identity, and the definition of $Q$, the left hand side of (\ref{property}) can be written in terms of the real non-singular matrix $R$ defined to be $R=\Delta(\bx,0)^\dagger \Delta(\bx,0)$.  After several cancellations this leads to
\begin{multline}
e_1^\dagger\big( Q(\bx,0)Q(\bx,\mu)-Q(\bx,-\mu)Q(\bx,0)\big)e_1 = \\
\mu L R^{-1} (M-\bx 1_N)\big(R+\mu^21_N\big)^{-1} L^\dagger 
-\mu L \big(R+\mu^2 1_N\big)^{-1} (M-\bx 1_N)R^{-1} L^\dagger.
\end{multline}
Using the identity $(R+\mu^2 1_N)^{-1}=R^{-1}(1_N+\mu^2 R^{-1})^{-1}$, this can be rearranged to
\begin{multline}
e_1^\dagger\big( Q(\bx,0)Q(\bx,\mu)-Q(\bx,-\mu)Q(\bx,0)\big)e_1 = \\
\mu L R^{-1}\Big[M-\bx 1_N,\big(1_N+\mu^2 R^{-1}\big)^{-1}\Big]
R^{-1} L^\dagger.\label{eq4}
\end{multline}
The commutator of $\big(1_N+\mu^2 R^{-1}\big)^{-1}$ with $\bx 1_N$ vanishes because the former is a real matrix and the latter is a multiple of the identity matrix.  By (\ref{axial}), the commutator with $M$ can be written in the form
\be
\Big[M,\big(1_N+\mu^2 R^{-1}\big)^{-1}\Big]=C_1i+C_2j.
\ee
Moreover, the real matrices $C_1,C_2$ are antisymmetric, because $M$ and $R$ are symmetric.  The right hand side of (\ref{eq4}) can be rewritten as
\be
\mu LR^{-1}(C_1i+C_2j)R^{-1}L^\dagger = \mu \mbox{Tr}\, R^{-1}L^tLR^{-1}(C_1i+C_2j)
\ee
because $k$ anticommutes with $i$ and $j$ and $L$ is written in terms of $1$ and $k$.  This expression vanishes because it is the trace of a product of a symmetric matrix and an antisymmetric matrix, and so (\ref{axial}) implies the property (\ref{property}).

This property, which is a kind of centring condition in the $x_4$ direction, in turn implies that the Skyrmion is rational.  The proof of this is as follows. As $Q(\bx,x_4)$ projects onto the kernel of $\Delta(\bx,x_4)$, which is spanned by the unit vector $\Psi(\bx,x_4)$, then
\be
\Psi(\bx,x_4)=\frac{Q(\bx,x_4)e_1}{|Q(\bx,x_4)e_1|},
\ee
therefore
\be
Q(\bx,x_4)=\Psi(\bx,x_4)\Psi(\bx,x_4)^\dagger
=\frac{Q(\bx,x_4)e_1e_1^\dagger Q(\bx,x_4)}{|Q(\bx,x_4)e_1|^2}.
\label{split}
\ee
Using this result,
\bea
&&e_1^\dagger Q(\bx,-\mu)Q(\bx,0)Q(\bx,\mu)e_1
=\frac{e_1^\dagger Q(\bx,-\mu) Q(\bx,0)e_1e_1^\dagger Q(\bx,0) Q(\bx,\mu)e_1}
{|Q(\bx,0)e_1|^2}\\
&=&
\frac{e_1^\dagger Q(\bx,0) Q(\bx,\mu)e_1e_1^\dagger Q(\bx,0) Q(\bx,\mu)e_1}
     {|Q(\bx,0)e_1|^2}
     =
\bigg(\frac{e_1^\dagger Q(\bx,0) Q(\bx,\mu)e_1}{|Q(\bx,0)e_1|}\bigg)^2,
\eea
where property (\ref{property}) has been used to obtain the first expression in the final line.
This shows that the numerator in (\ref{scheme5}) is the square of a quaternion, hence dividing by the modulus indeed yields a rational rather than an algebraic expression.

\subsection{Charge one}\quad
Consider $N=1$ with $L=\lambda$ and $M=0$, then (\ref{scheme5}) gives a rational hedgehog Skyrmion
\be
U=
\frac{r^{2}(r^{2}+\lambda^{2}+\mu^{2})^2-\lambda^{4} \mu^{2}+2i\lambda^2\mu (r^{2}+\lambda^{2}+\mu^{2})\bx\cdot\boldsymbol{\tau}}{\big((r^{2}+\lambda^{2})^2+\mu^{2}r^{2}\big) \left(r^2+\mu^{2}\right)}.
\ee
Writing this Skyrmion in the hedgehog form
(\ref{hedgehog}) gives the profile function 
\be
f(r)=\tan^{-1}\bigg(
\frac{2 \lambda^{2} \mu  r \left(r^{2}+\lambda^{2}+\mu^{2}\right)}{r^{2} \left(r^{2}+\lambda^{2}+\mu^{2}\right)^{2}-\lambda^{4} \mu^{2}}
\bigg).
\label{newprofile5}
\ee
Minimizing the spherical energy (\ref{sphericalenergy}) with this
profile function yields a value $E=1.236$, obtained for $\lambda=2.03, \ \mu=1.43.$
This energy is only $0.3\%$ above the true Skyrmion energy $E=1.232$, and 
is lower than the value $E=1.243$ obtained by the Atiyah-Manton profile function
(\ref{amprofile}) with the optimal instanton scale $\lambda=1.45.$
Note that the profile function (\ref{newprofile5}) decays like $r^{-3}$.  In contrast, the Atiyah-Manton profile function (\ref{amprofile}) and the true profile function of the Skyrmion both decay like $r^{-2}$.

\begin{figure}[!ht]\begin{center}
    \includegraphics[width=0.5\columnwidth]{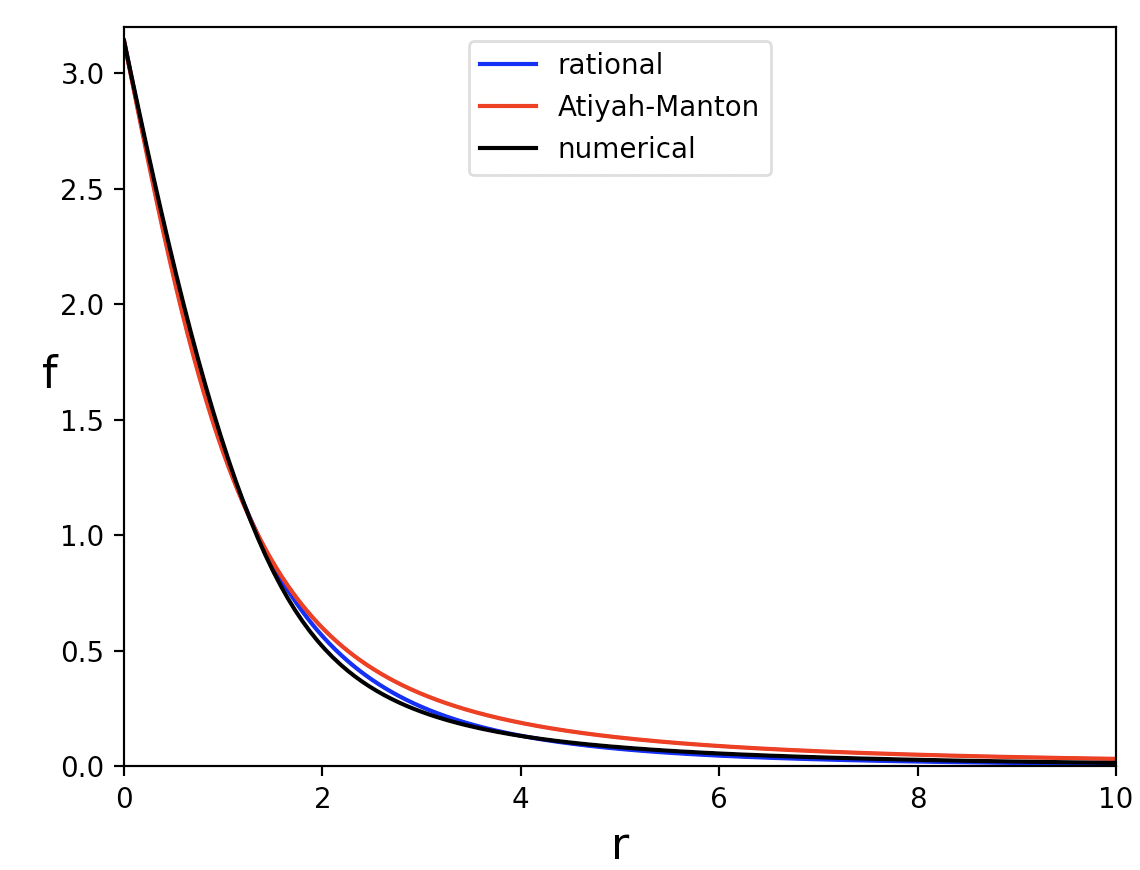}
    \caption{
      The profile function of the charge one rational Skyrmion (blue), the Atiyah-Manton profile function (red), and the true profile function (black) obtained numerically.}
    \label{fig:profiles}\end{center}\end{figure}

Fig.\ref{fig:profiles} displays the rational Skyrmion profile function (\ref{newprofile5}) as the blue curve, and for comparison, the Atiyah-Manton profile function (\ref{amprofile}) as the red curve, and the true profile function obtained numerically as the black curve. This confirms that the charge one rational Skyrmion is an excellent approximation to the true Skyrmion and is even better than the Atiyah-Manton approximation. 

Setting $\lambda=2, \ \mu=\sqrt{2}$, which are close to the numerically determined values, does not change the energy at the level of the precision given above. The rational hedgehog Skyrmion then simplifies to
\be
U=
\frac{r^{6}+12 r^{4}+36 r^{2}-32+8\sqrt{2}i (r^{2}+6)\bx\cdot\boldsymbol{\tau}}{\left(r^{2}+8\right) \left(r^{2}+2\right)^{2}}.
\ee

\subsection{Charge two}\quad
The ADHM data for an axially symmetric charge two Skyrmion is
\be
\widehat M = \begin{pmatrix} L\\ M \end{pmatrix}
=\frac{\lambda}{2}
\begin{pmatrix} 
\sqrt{2} & \sqrt{2}k\\
i&j\\
j&-i
\end{pmatrix},
\label{Mhat2}
\ee
from which the formula (\ref{scheme5}) produces a rational Skyrmion that will be written in the form
\be
U=\frac{{\widetilde\sigma+i\widetilde{\boldsymbol{\pi}}\cdot\boldsymbol{\tau}}}{\sqrt{\widetilde\sigma^2+\widetilde\pi_1^2+\widetilde\pi_2^2++\widetilde\pi_3^2}}.
\label{unnorm}
\ee
Motivated by the parameter values from the $N=1$ case, setting $\lambda=2$ and
$\mu=\sqrt{2N}=2$ simplifies the expression for the Skyrmion, which when 
written in terms of $x_1,x_2,x_3, r^2$ and $\rho^2=x_1^2+x_2^2$ is
\bea
 \widetilde\pi_1&=&64 \left(x_{1}^{2}-x_2^2\right) \left(r^{2}+4\right) \left((r^{2}+8)^2-4\rho^2\right)
 \nonumber\\
 \widetilde\pi_2&=&128x_{1}x_2 \left(r^{2}+4\right) \left((r^{2}+8)^2-4\rho^2\right)
  \\
  \widetilde\pi_3&=&
  16 x_{3} \left(r^{4}+12 r^{2}+4 \rho^{2}+32\right)\left((r^{2}+8)^2-4\rho^2\right)
  \nonumber\\
  \widetilde\sigma&=&
r^{12}+36 r^{10}+\left(-12 \rho^{2}+512\right) r^{8}+\left(-288 \rho^{2}+3520\right) r^{6}+\left(48 \rho^{4}-2496 \rho^{2}+11008\right) r^{4}\nonumber\\
&&\qquad+\left(576 \rho^{4}-9728 \rho^{2}+8192\right) r^{2}-64 \rho^{6}+1792 \rho^{4}-16384 (\rho^{2}+1).\nonumber
\eea
The ADHM data (\ref{Mhat2}) satisfies the condition (\ref{axial}), therefore the Skyrmion is rational. An explicit calculation confirms that
$\Xi=\sqrt{\widetilde\sigma^2+\widetilde\pi_1^2+\widetilde\pi_2^2+\widetilde\pi_3^2}$ is given by
  \be
\Xi= \left(r^{8}+24 r^{6}+8\left(24- \rho^{2}\right) r^{4}+96\left(6 - \rho^{2}\right) r^{2}+16(\rho^{4}-28 \rho^{2}+32)\right) \left(r^{4}+12 r^{2}-4 \rho^{2}+32\right).\nonumber  
\ee
The energy is $E=2.418$, which is very close to the energy $E=2.416$ of the rational map approximation, with both being $2.5\%$ larger than the true minimal energy $E=2.358.$ For comparison, the energy $E=2.384$ of the numerically computed Atiyah-Manton axial charge two Skyrmion has an error of only $1.1\%.$ 
Calculating the rational Skyrmion for general $\lambda$ and $\mu$ reveals that the energy is minimized for the parameter values $\lambda=1.99, \ \mu=2.06$,
which are close enough to the chosen values $\lambda=\mu=2$ that the energy is unchanged at the level of the precision given.

A family of charge two Skyrmions in the attractive channel is obtained from the ADHM data \cite{MS}
\be
\widehat M=\frac{\lambda}{2}
\begin{pmatrix} 
\sqrt{2(1-a^2)} & \sqrt{2(1-a^2)}k\\
(1+a)i&(1-a)j\\
(1-a)j&-(1+a)i
\end{pmatrix} \,,
\label{adhm21}
\ee
where $a\in(-1,1)$ controls the separation of the pair of Skyrmions, with the axial case recovered when $a=0$, and the separation tending to infinity as $|a|\to 1.$
This ADHM data satisfies the condition (\ref{axial}) and,
using the notation (\ref{unnorm}), yields the family of rational Skyrmions
\be\begin{split}
&\widetilde\pi_1=2\left( 1-a^{2}\right) \mu  \,\lambda^{3}H_1H_2,\quad
\widetilde\pi_2=4\left(1- a^{2}\right) \mu  \,\lambda^{3}x_1x_2H_1H_3,\quad
\widetilde\pi_3=2\left(1- a^{2}\right) \mu  \,\lambda^{2}x_3H_1H_4,\\
&\widetilde\sigma=
-\frac{\mu^{2}}{4} \left(1-a^2\right)^{5} \lambda^{10}
-\frac{\mu^{2}}{4}\left(1 -a^2\right)^{4}  \left(2\mu^{2}+2\rho^{2}+{3 H_{3}}\right) \lambda^{8}
-\frac{(1-a^2)}{2}H_{1}^{2} \left(\mu^{2}+H_{3}\right) \lambda^{2}
\\
&-\frac{\mu^{2}}{4}\left(1 -a^2\right)^{3} \left(\mu^{4}+\mu^2\left(2 \rho^{2}+4 H_{3}\right) +2 \rho^{2} H_{3}+3 H_{3}^{2}-2 H_{1}\right)  \lambda^{6}
-\mu^{2} H_{1}^{2} H_{3}+H_{1}^{3}\\
&-\frac{\left(1 -a^2\right)^{2}}{4} \left(\mu^{6} H_{3}+\left(2 H_{3}^{2}-2 H_{1}\right) \mu^{4}+\left(-4 \rho^{2} H_{1}+H_{3}^{3}-2 H_{1} H_{3}\right) \mu^{2}+2 H_{1}^{2}\right) \lambda^{4},
\end{split}\ee
where the following functions have been introduced for notational convenience
\bea
H_1&=&\lambda^{4}+\left(2 \left(x_{1}^{2}-x_{2}^{2}\right) a +2 r^{2}+2 \mu^{2}-(1+a^2)\rho^{2}\right) \lambda^{2}+\left(\mu^{2}+r^{2}\right)^{2},
\nonumber\\
H_2&=&\left(x_{1}^{2}-x_{2}^{2}\right) \left(\lambda^{2}(1-a^2)+\mu^{2}+2 r^{2}\right)+\left(\lambda^{4}+\left(\mu^{2}+2 r^{2}\right) \left(\lambda^{2}-\rho^{2}\right)+r^{4}+r^{2} \mu^{2}\right) a,
\nonumber\\
H_3&=&\lambda^{2} \left(a^{2}+1\right)+2 r^{2}+\mu^{2},
\nonumber\\
H_4&=&\lambda^{4}+\left(a^{2} \rho^{2}-2 \left(x_{1}^{2}-x_{2}^{2}\right) a +2 r^{2}+\mu^{2}+\rho^{2}\right) \lambda^{2}+r^{4}+r^{2} \mu^{2}.
\eea
This family is invariant under the transformation $(a,x_1,x_2,x_3)\mapsto (-a,x_2,-x_1,x_3)$ together with the compensating isospin rotation $(\sigma,\pi_1,\pi_2,\pi_3)\mapsto (\sigma,-\pi_1,-\pi_2,\pi_3).$ Therefore, to study separated Skyrmions it is enough to restrict to the case $a\in(0,1)$, where the pair of Skyrmions are positioned on the $x_1$-axis.
 \begin{figure}[!h]\begin{center}
    \includegraphics[width=1.0\columnwidth]{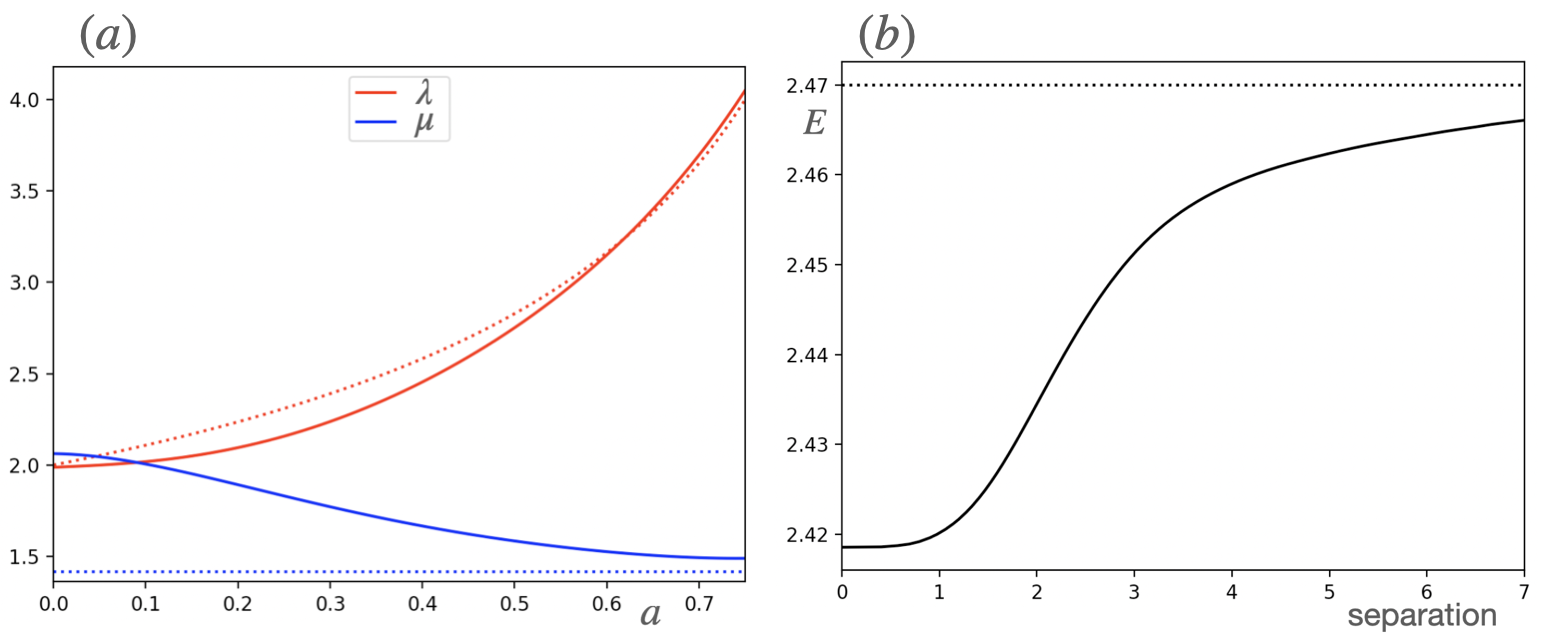}
    \caption{(a) The parameters $\lambda$ and $\mu$ as a function of $a$, together with their asymptotic approximations (dotted curves), valid for $a\approx 1$. (b) The energy as a function of the Skyrmion separation.}
    \label{fig:ch2}\end{center}\end{figure}

 To consider the well-separated limit $a\approx 1$, write
$
\lambda={\lambda_1}/{\sqrt{1-a}},
$
where $\lambda_1$ is independent of $a$. In this limit the ADHM data (\ref{adhm21}) simplifies to
\be
\widehat M=
\begin{pmatrix} 
\lambda_1 & \lambda_1k\\
i\lambda_1/\sqrt{1-a}&0\\
0&-i\lambda_1/\sqrt{1-a}
\end{pmatrix} +O(\sqrt{1-a})\,,
\label{adhm21lim}
\ee
and describes well-separated instantons \cite{CWS} because the diagonal entries of $M$ are much larger than any of the other entries of $\widehat M$. This generates a pair of well-separated Skyrmions positioned on the $x_1$-axis at $x_1=\pm \lambda_1/\sqrt{1-a}$, with scale $\lambda_1.$ From the earlier analysis of the charge one case, in this limit the energy minimizing parameters are $\lambda_1\approx2,\ \mu\approx\sqrt{2}.$
 
Fig.\ref{fig:ch2}(a) displays the energy minimizing parameters $\lambda$ and $\mu$, as a function of $a$. The dotted curve is the asymptotic approximation
$\lambda\approx2/\sqrt{1-a}$, that is derived for $a\approx 1$, but turns out to be a reasonable approximation for all $a\in[0,1).$ The dotted line is the asymptotic approximation $\mu\approx \sqrt{2}$, which is only valid for $a\approx 1.$
Fig.\ref{fig:ch2}(b) plots the energy as a function of the separation between the two Skyrmions, where the Skyrmion positions are defined to be the points at which $U=-1.$ The dotted line is the asymptotic value at infinite separation, given by twice the energy of a single rational Skyrmion. This graph clearly illustrates the attractive force between the pair of Skyrmions, and is the first time that such a calculation has been performed using an explicit Skyrme field that is applicable for all separations.
 
\begin{figure}[!ht]\begin{center}
    \includegraphics[width=1.0\columnwidth]{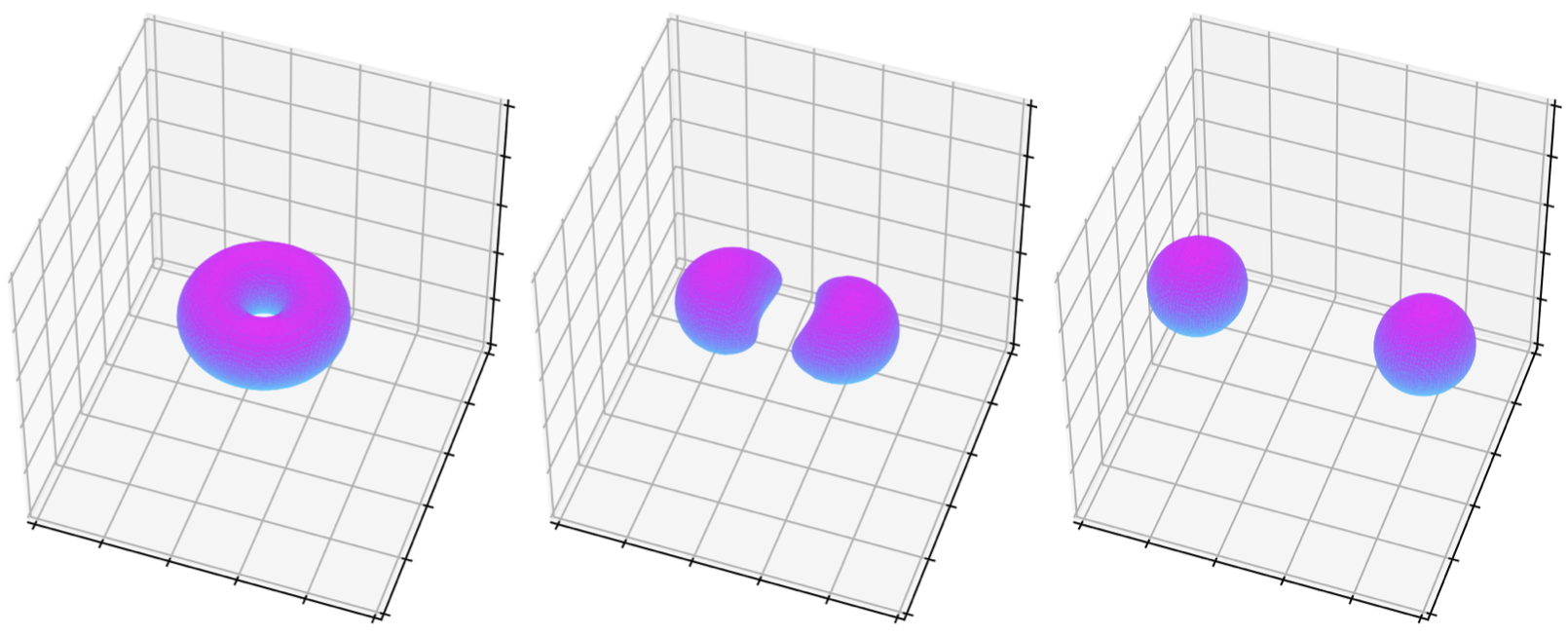}
    \caption{
Baryon density isosurfaces for charge two rational Skyrmions with $a=0,0.3,0.6.$}
    \label{fig:bden}\end{center}\end{figure}

Isosurfaces on which the baryon density (the intergrand in (\ref{baryon})) is equal to 0.05 are displayed in Fig.\ref{fig:bden} for the values $a=0,0.3,0.6.$ This shows how the axial charge two Skyrmion separates into a pair of charge one Skyrmions as the parameter $a$ is increased from zero.

 There is a spherically symmetric charge two Skyrmion that is a saddle point of the energy. It is given by the hedgehog form (\ref{hedgehog})
with a profile function $f(r)$ that satisfies the boundary conditions $f(0)=2\pi$ and $f(\infty)=0.$ The value of the energy is $E=3.667$, which is substantially larger than twice the energy of a single Skyrmion. A rational Skyrmion approximation to this saddle point is obtained from the ADHM data
 \be \widehat M=\begin{pmatrix} \lambda & \,\lambda \\ \nu & 0 \\ \,0 & -\nu \end{pmatrix},
 \ee
 where $\lambda$ and $\nu$ are real positive parameters.
Note that this ADHM data does not have axial symmetry in the $(x_3,x_4)$-plane, with (\ref{axial}) not satisfied. However, a direct calculation confirms that (\ref{property}) is satisfied for this data and therefore it yields a rational Skyrmion. This example shows that (\ref{axial}) is a sufficient but not necessary condition for the Skyrmion to be rational. A full investigation of which ADHM data lead to rational Skyrmions is beyond the scope of this paper.

 The energy is minimized  for the parameter values, $\lambda=3.50,\, \nu=2.94,\, \mu=3.94,$ giving an energy
 $E=3.820$, which is $4.2\%$ above the true value. Setting the parameter values
 to $\lambda=7/2,\, \nu=3,\, \mu=4,$ which are close enough to the numerical values to have the same energy to the given precision, produces the profile function
 \be
 f(r)=\tan^{-1}\bigg(
   \frac{392 r \left(2 r^{8}+217 r^{6}+6839 r^{4}+64395 r^{2}-55125\right)}{4 r^{12}+668 r^{10}+38545 r^{8}+842540 r^{6}+4439626 r^{4}-35758240 r^{2}+4862025}
 \bigg).
 \ee
 The Atiyah-Manton approximation to the spherical charge two Skyrmion has an energy $E=3.711$, which is only $1.2\%$ above the true value. The reason that the rational Skyrmion has a relatively large error for this example is that the instanton has a more complicated $x_4$ dependence than the earlier examples, where the instanton was localized around $x_4=0.$ Sampling the ADHM data at only three interior points, $x_4=0,\pm \mu$, is therefore less accurate in this situation. 
   
\section{Algebraic Skyrmions}\quad
In the previous section it has been demonstrated that a good rational approximation to Skyrmions can be generated using only five points in the formula (\ref{anyp}). Fewer points lead to simpler formulae, so it is tempting to investigate versions of the scheme with even fewer points.
The simplest scheme is a 3-point formula that uses the points $t_1,t_2,t_3=-\infty,0,\infty$, but this turns out to be trivial, as it generates only the vacuum Skyrme field $U=1.$  This is easily shown as follows,
\be
U(\bx)=\frac{e_1^\dagger Q(\bx,0) e_1}{|e_1^\dagger Q(\bx,0) e_1|}
=\frac{(e_1^\dagger Q(\bx,0) e_1)^\dagger}{|e_1^\dagger Q(\bx,0) e_1|}
=U(\bx)^\dagger,
\ee
which implies that $U=1$.

The 4-point scheme with points $t_1,t_2,t_3,t_4=-\infty,-\mu,\mu,\infty$, is
\be
U(\bx)=\frac{e_1^\dagger Q(\bx,-\mu) Q(\bx,\mu) e_1}{|e_1^\dagger Q(\bx,-\mu) Q(\bx,\mu) e_1|},
\label{scheme4}
\ee
and does generate Skyrmions, although they are algebraic rather than rational.

Taking the $N=1$ data $L=\lambda$ and $M=0$ gives a hedgehog Skyrmion
\be
U=\frac{(r^2+\mu^2)^2+\lambda^2(r^2-\mu^2)+2i\lambda^2\mu \bx\cdot\boldsymbol{\tau}}
{(r^2+\mu^2)\sqrt{(r^2+\mu^2+\lambda^2)^2-4\mu^2\lambda^2}},
\ee
with charge one, providing $\mu\in(0,\lambda).$ The corresponding profile function is
\be
f(r)=\tan^{-1}\bigg(
\frac{2\lambda^2\mu r}{(r^2+\mu^2)^2+\lambda^2(r^2-\mu^2)}
\bigg),
\label{profilescheme4}
\ee
with a minimal energy $E=1.240$ obtained when $\lambda=2\mu=2.48.$ This energy is very slightly lower than the energy of the Atiyah-Manton profile, but it is above that of the rational Skyrmion.

Using the ADHM data (\ref{Mhat2}) yields the axially symmetric charge two algebraic Skyrmion
\begin{eqnarray}
\widetilde\pi_1&=&
2 \lambda^{3}\mu \left(2\mu^{2}+{\lambda^{2}}+2r^{2}\right) \left(x_{1}^{2}-x_{2}^{2}\right),\nonumber\\
\widetilde\pi_2&=&
4\lambda^{3}\mu  \left(2\mu^{2}+{\lambda^{2}}+2r^{2}\right) x_{1}  x_{2},\nonumber\\ 
\widetilde\pi_3&=&
2 \lambda^{2} \mu  \left
((x_{3}^{2}+\lambda^2)^2+2 (\mu^{2}+ \rho^{2}) x_{3}^{2}
+\left(2 \mu^{2}+3 \rho^{2}\right) \lambda^{2}+\left(\mu^{2}+\rho^{2}\right)^{2}\right) x_{3},\nonumber\\
\widetilde\sigma&=&
\left(x_{3}^{2}-\mu^{2}\right) \lambda^{6}+\left(\left(-3 \rho^{2}+2 x_{3}^{2}\right) \mu^{2}+\rho^{4}+3 \rho^{2} x_{3}^{2}+3 x_{3}^{4}-\mu^{4}\right) \lambda^{4}\nonumber\\
&&+\left(\mu^{2}+\rho^{2}+3 x_{3}^{2}\right) \left(\mu^{2}+r^2\right)^{2} \lambda^{2}+\left(\mu^{2}+r^2\right)^{4},
\end{eqnarray}
which again requires that $\mu\in(0,\lambda).$
The energy is minimized for the parameter values $\lambda=2.34, \ \mu=1.28$, giving $E=2.468$, which is $4.7\%$ larger than the true minimal energy, so almost twice the $2.5\%$ error of the corresponding rational Skyrmion.

\section{Conclusion}\quad
A new method has been introduced that provides a simple formula to produce explicit rational approximations to Skyrmions from ADHM instanton data. The method is an improvement on the related Atiyah-Manton procedure, that obtains approximate Skyrmions as instanton holonomies, in that the Skyrme field is obtained in closed form using only linear algebra, and does not require the solution of any differential equations. Despite this simplification, the new method has a similar level of accuracy to the Atiyah-Manton approach, and is a better approximation for a single Skyrmion.

The moduli space of charge $N$ instantons induces a moduli space of charge $N$ rational Skyrmions. As an illustration, rational Skyrmions have been used to provide the first example of a family of explicit charge two Skyrme fields that includes a pair of well-separated Skyrmions, the axially symmetric charge two Skyrmion, and all intermediate Skyrmions in the attractive channel with any separation. Moduli spaces of rational Skyrmions should prove useful in the quantization of Skyrmions, with simplifications resulting from the availability of explicit Skyrme fields.

Finally, there are several known examples of ADHM data that are appropriate for studying minimal energy Skyrmions with charges greater than two, including families that contain separation parameters to deform the minimal energy Skyrmion into individual Skyrmions or clusters in a particularly symmetric way. All of these examples are related to hyperbolic monopoles and the ADHM data has an axial symmetry that guarantees a rational Skyrmion. It might be interesting to investigate the details.

\end{document}